\documentclass[manuscript]{aastex}
\usepackage{natbib}
\usepackage{epsfig}
\slugcomment{}

\def\beq{\begin{eqnarray}}
\def\eeq{\end{eqnarray}}

\begin{document}

\title{Self-gravity in neutrino-dominated accretion disks}

\author{Tong Liu$^{1,2,3}$, Xiao-Fei Yu$^{1}$, Wei-Min Gu$^{1}$, and Ju-Fu Lu$^{1}$}
\altaffiltext{1}{Department of Astronomy and Institute of Theoretical Physics and Astrophysics, Xiamen University, Xiamen, Fujian 361005, China; tongliu@xmu.edu.cn}
\altaffiltext{2}{Key Laboratory for the Structure and Evolution of Celestial Objects, Chinese Academy of Sciences, Kunming, Yunnan 650011, China}
\altaffiltext{3}{State Key Laboratory of Theoretical Physics, Institute of Theoretical Physics, Chinese Academy of Sciences, Beijing 100190, China}

\begin{abstract}
We present the effects of the self-gravity on the vertical structure and neutrino luminosity of the neutrino-dominated accretion disks in cylindrical coordinates. It is found that significant changes of the structure appear in the outer region of the disk, especially for high accretion rates (e.g., $\ga 1 M_\odot~\rm s^{-1}$), and thus causes the slight increase of the neutrino luminosity. Furthermore, the gravitational instability of the disk is reviewed by the vertical distribution of the Toomre parameter \citep{Toomre1964}, which may account for the late-time flares in gamma-ray bursts and the extended emission in short-duration gamma-ray bursts.
\end{abstract}


\section{Introduction}

Self-gravity has been widely investigated in accretion systems \citep[see, e.g.,][]{Paczynski1978a,Paczynski1978b,Abramowicz1984,Goodman1988}. It is easy to find that the self-gravity is important and its resulting local instabilities may develop if the mass density becomes comparable to $M/R^3$, where $M$ and $R$ are the mass of the central object and the disk radius, respectively. The effects of the self-gravity have been considered in some interpretations of astrophysical processes. The self-gravity constraint limits the scale and angular momentum of the disk to affect the evolution of black hole mass and spin in active galactic nuclei \citep[e.g.,][]{King2008,Hopkins2010}. It also influences the star formation in galaxies and the formation of protostars and protostellar disks \citep[e.g.,][]{Goodman2003,McKee2007,Rice2010}.

Gamma-ray bursts (GRBs) can be sorted into two classes, i.e., short-duration and long-duration GRBs \citep{Kouveliotou1993}. Whatever types they are, the central engine of GRBs is usually considered as the system consisting of a hyperaccreting spinning stellar-mass black hole with the mass accretion rate in the range of $0.003-10$ $M_\odot ~\rm s^{-1}$, surrounded by a geometrically and optically thick disk with the high density and temperature, especially for the inner region ($\rho \sim 10^{10}~\rm g~cm^{-3}$, $T \sim 10^{10}~\rm K$), namely neutrino-dominated accretion flow (NDAF). The annihilation of neutrinos escaping from the disk surface can power GRBs. This model has been studied in the past decades \citep[see, e.g.,][]{Popham1999,Narayan2001,Di Matteo2002,Kohri2002,Kohri2005,Lee2005,Gu2006,Chen2007,Kawanaka2007,Kawanaka2012,Liu2007,Liu2008,Liu2010a,Liu2010b,Liu2012a,Liu2012b,Liu2013,Liu2012,Pan2012,Xue2013}. The properties of such a disk model were first worked out in details by \citet{Popham1999}. The detailed microphysics and the strict hydrodynamics and thermodynamics have been considered in some subsequent works, such as one dimensional cases \citep[e.g.,][]{Di Matteo2002,Kohri2005,Chen2007,Janiuk2007,Kawanaka2007,Liu2007,Li2013,Xue2013} and two dimensional cases \citep[e.g.,][]{Barkov2011,Janiuk2013}. The dynamical features of the NDAF, such as the jet precession triggered by the system of the black hole and disk \citep[e.g.,][]{Reynoso2006,Liu2010b,Sun2012,Hou2014a,Hou2014b} can explain the quasi-periodic structure in a wide variety of observed light profiles of GRBs, and the shapes of the light curves (particularly for those with a fast rise and exponential decay or an approximate symmetry), and components features, i.e., nucleosynthesis near the surface of the disk \citep[e.g.,][]{Fujimoto2004,Surman2004,Banerjee2013,Liu2013,Xue2013}, may provide a clue to understand the bumps in the optical light curve
of core-collapse supernovae and the strong Fe K$\alpha$ emission lines in some GRB observaions \citep[e.g.,][]{Piro2000}.

Due to the high mass density of the NDAF, we argue that the self-gravity effect may be important to the structure of the disk, and the neutrino luminosity further. Moreover, \citet{Perna2006} studied the gravitational instability in the outer parts of the hyperaccretion disk in the center of GRBs, which may result in actual fragmentation of the disk to produce energetic flares of GRBs. The vertical structure of the optically thick disk, including NDAF was investigated \citep[e.g.,][]{Gu2007,Liu2008,Gu2009,Jiao2009,Liu2010a}, but the effects of the self-gravity and the gravitational instabilities have not been considered in these works.

In this paper, we focus on the effects of the self-gravity on the vertical structure and the neutrino luminosity of the NDAF for varying accretion rates. In Section 2, the basic equations in the vertical direction and the Toomre parameter \citep{Toomre1964} are introduced. The numerical results of the disk and the vertical distribution of the Toomre parameter are presented in Section 3. Conclusions and discussion are made in Section 4.

\section{Physical model}
\subsection{Basic equations}
NDAF is one of the geometrically thick disks \citep[e.g.,][]{Popham1999,Liu2007,Liu2008,Liu2010a} as well as the slim disk \citep{Abramowicz1988}, so its hydrodynamics and thermodynamics are expected to be similar to those of slim disks. The angular velocity $\Omega$ is approximately Keplerian, i.e., $\Omega=\Omega_{\rm K}$, Thus we can write the continuity, angular momentum, and energy equations of the NDAF in cylindrical coordinates following \citet{Gu2007} and \citet{Liu2008}:
\beq
{\dot M} =-2 \pi R \Sigma v_R = {\rm constant},
\eeq
\beq
{\dot M} (\Omega_{\rm K} R^2-j) = 2 \pi \alpha R^2 \Pi,
\eeq
\beq
Q_{\rm vis} = Q_{\rm adv} + Q_{\nu},
\eeq
where $\dot M$ is the mass accretion rate, $v_R$ is the radial velocity, $\Omega_{\rm K}=(GM/R)^{1/2}/(R-R_{\rm g})$ is the Keplerian angular velocity, $G$ is the gravitation constant, $R_{\rm g}=2GM/c^2$ is the Schwarzschild radius, $j=1.8c R_{\rm g}$ is an integration constant representing the specific angular momentum accreted by the black hole, and $\alpha$ is the Shakura-Sunyaev viscosity parameter.

Furthermore, $\Sigma$ and $\Pi$ are the surface density and vertically integrated pressure, respectively, which can be defined as
\beq
\Sigma =2 \int_{0}^{\infty} {\rho} {\rm d} z,
\eeq
\beq
\Pi =2 \int_{0}^{\infty} {p} {\rm d} z,
\eeq
where $\rho$ and $p$ are the mass density and the pressure of the disk, respectively, and the sound speed is further defined as $c_{\rm s} = (\Pi/\Sigma)^{1 / 2}$. Here we define the half thickness of the disk $H=\Sigma/2 \rho_0$, where $\rho_0$ is the mass density on the equatorial plane.

The viscous heating rate is
\beq
Q_{\rm vis} =\frac{1}{2 \pi}{\dot M} \Omega_{\rm K} ^2 f g,
\eeq
where $f = 1 - j/\Omega_{\rm K} R^2$, and $g = - {\rm d ln} \Omega_{\rm K}/{\rm d ln} R$. The advective cooling rate is
\beq
Q_{\rm adv} =\frac{1}{2\pi} \frac{\xi {\dot M} {c_{\rm s}}^2}{R^2},
\eeq
with $\xi=3/2$ being a dimensionless quantity of the order of unity \citep[e.g.,][]{Kato2008,Liu2008}. The neutrino cooling is expressed by a bridging formula \citep[e.g.,][]{Di Matteo2002,Kohri2005,Liu2007} that is valid in both the neutrino optically thin and thick regimes of the disk:
\beq
Q_{\nu}=\sum_{i} \frac{(7/8) {\sigma} T^4}{(3/4)[\tau_{{\nu}_i}/2+1/ \sqrt{3}+1/(3 \tau_{a,{\nu}_i})]},
\eeq
where $\sigma$ is the Stefan-Boltzmann constant, $T$ is the temperature, and $\tau_{{\nu}_i}$ is the total optical depth for neutrinos, including the absorption optical depth $\tau_{a,{\nu}_i}$ and scattering optical depth $\tau_{s,\nu_i}$,
\beq
\tau_{\nu_i}=\tau_{a,\nu_i}+\tau_{s,\nu_i},
\eeq
where the subscript $i$ runs over the three species of neutrinos $\nu_{\rm e}$ , $\nu_\mu$ , and $\nu_\tau$. Here we ignore the production and distribution of the heavy elements \citep[e.g.,][]{Di Matteo2002,Gu2006,Liu2007,Liu2008}. The main absorption processes includes the electron-positron pair annihilation and Urca processes \citep[e.g.,][]{Narayan2001,Di Matteo2002,Liu2007}, the corresponding optical depths can be written as
\beq
\tau_{a,\nu_i}=2.5\times10^{-7}T_{11}^5 H,
\eeq
\beq
\tau_{a,\nu_e2}=2.5\times10^{-7}T_{11}^2 X_{\rm nuc} \rho_{10} H,
\eeq
where $T_{11}=T/10^{11} \rm K$, $\rho_{10}=\rho/10^{10} \rm g~cm^{-3}$, and $X_{\rm nuc}$ is the mass fraction of free nucleons approximately given by \citep[e.g.,][]{Liu2007}
\beq
X_{\rm nuc}={\rm min}\{1,~295.5 {\rho_{10}}^{-3/4} {T_{11}}^{9/8} {\rm exp}(-0.8209/T_{11})\}.
\eeq
The total optical depth of scattering by nucleons can be given by
\beq
\tau_{s,\nu_i}=2.7\times10^{-7}T_{11}^2 \rho_{10} H.
\eeq

The equation of state is  written as \citep[e.g.,][]{Di Matteo2002,Liu2008}
\beq
p = p_{\rm gas} + p_{\rm rad} + p_{\rm e} + p_{\nu}.
\eeq
The gas pressure from nucleons, radiation pressure of photons, degeneracy pressure of electrons, and radiation pressure of neutrinos can be expressed as $p_{\rm gas}=(\rho k_{\rm B} T/m_u)(1+3 X_{\rm nuc})/4$, $p_{\rm rad}=aT^4/3$, $p_{\rm e}=(2\pi h c/3)[3\rho/(16\pi m_u)]^{4/3}$, and $p_\nu=u_{\rm \nu}/3$, respectively, where $k_{\rm B}$ is the Boltzmann constant, $m_u$ is the mean mass of a nucleon, $h$ is the Planck constant, $a$ is the radiation density constant, and the energy density of neutrinos $u_{\rm \nu}$ is \citep[e.g.,][]{Kohri2005,Liu2007,Liu2008}
\beq
u_{\rm \nu}=\sum_{i} \frac{(7/8)a T^4 (\tau_{{\nu}_i}/2+1/ \sqrt{3})} {\tau_{{\nu}_i}/2+1/ \sqrt{3}+1/(3 \tau_{a,{\nu}_i})}.
\eeq
Moreover, we assume a polytropic relation in the vertical direction, $p=K\rho^{4/3}$, where $K$ is a constant \citep{Liu2008,Liu2010a}.

If the self-gravity is considered in the NDAF model, the vertical equilibrium equation should be rewritten. We want to compare the cases considered with and without the self-gravity. The vertical equilibrium equation is given by the following three forms.
\begin{enumerate}

\item
Case I - The form with self-gravity

As well as \citet{Paczynski1978a,Paczynski1978b}, we define the surface density of the disk in the range from the equatorial plane to a certain height $z$ ($z\leq H$),
\beq
\Sigma_z = \int_{0}^{z} {\rho} {\rm d} z'.
\eeq
If it varies slowly with radius, a new term $4\pi G \Sigma_z$ should be added in the vertical equilibrium equation, which represents the self-gravity of the disk. The equation can be written as
\beq
4\pi G \Sigma_z + \frac{\partial \Psi}{\partial z} + \frac{1}{\rho} \frac{\partial p}{\partial z} =0,
\eeq
where $\Psi$ is the pseudo-Newtonian potential written by \citep{Paczynski1980}
\beq
\Psi = - \frac {G M}{\sqrt{R^2+z^2}-R_{\rm g}}.
\eeq

\item
Case II - The form without self-gravity

Following \citet{Gu2007} and \citet{Liu2008}, the vertical equilibrium equation without the self-gravity can be expressed as
\beq
\frac{\partial \Psi}{\partial z} + \frac{1}{\rho} \frac{\partial p}{\partial z} =0.
\eeq

\item
Case III - analytical form of Case I

In order to compare the cases with and without the self-gravity and indicate the vertical properties, we replace the terms of the vertical equilibrium equation in Case I [Equation (17)] with physical quantities on the equatorial plane of the disk (the subscript ``0''),
\beq
4 \pi G \rho_0 H +\Omega_{\rm K}^2 H - \frac{p_0}{\rho_0 H}=0,
\eeq
which can be regarded as a reference. Moreover, the vertically integrated pressure $\Pi$ is simplified as $2 p_0 H$.
\end{enumerate}

Finally, boundary conditions are required to numerically solve equations (2-19), which is suggested that $\rho$ and $p$ tend to be zero at the surface of the disk.

\subsection{Toomre parameter}

The matter of a differentially rotating disk is against gravitational collapse itself. The Toomre parameter can be used to measure the local gravitational stability of the accretion disks, which is expressed as
\beq
Q=\frac{c_{\rm s} \Omega_{\rm K}}{\pi G \Sigma_z},
\eeq
where $Q<1$ implies instability. This criterion is also widely used in researches on the star formation and protostellar disks and so on. If the effects of the self-gravity are considered in the vertical structure of NDAFs, the gravitational instability should be also reviewed in the framework.

\section{Numerical Results}

Our results for the vertical structure and neutrino luminosity of the NDAF are shown in Figures 1-6. In all these figures, the necessary constant parameters are fixed to their most typical values, that is, $\alpha=0.1$ and $M=3M_\odot$.

\subsection{Vertical structure}

Figure 1 shows the variations of the mass density on the equatorial plane $\rho_0$ with radius $R$ from 3$R_{\rm g}$ to 200$R_{\rm g}$ for $\dot{m}=0.1,~1,~10$ ($\dot M=\dot{m} M_{\odot}~\rm s^{-1}$). Case I, II, and III are described by the solid, dashed, and dotted lines, respectively. Figure 2 shows the variations of $\rho_0$ with accretion rate $\dot M$ from 0.1$M_{\odot}~\rm s^{-1}$ to 10$M_{\odot}~\rm s^{-1}$ for fixed radii 10$R_{\rm g}$ and 100$R_{\rm g}$. Obviously, the effects of the self-gravity mainly reflect in the outer region of the disk, especially for the high accretion rates.

Figures 3 and 4 show that the relative thickness of the disk $H/R$ as functions of the radius $R$ and accretion rate $\dot M$, respectively. We also find that the effects of the self-gravity mainly reflect in the outer region, especially for higher accretion rate. As shown in Figure 3(c), $H/R$ in Case I is less than that in Case II by nearly an order of magnitude at 200$R_{\rm g}$ for 10$M_{\odot}~\rm s^{-1}$. We also notice that the density and the thickness have the abnormal changes in the outer region of the disk for low accretion rates. For example, when the accretion rate is 0.1$M_{\odot}~\rm s^{-1}$, the abnormal behaviors exist in the region outside of 50$R_{\rm g}$. The reason is that the degeneracy pressure of electrons, replacing the gas pressure from nucleons, dominates in the outer region for low accretion rates, i.e. $0.1M_\odot~\rm s^{-1}$, thus causes the thickness $H$ larger and the density $\rho_0$ lower in Case I than those in Case II.

Contours of the Toomre parameter with the cylindrical coordinates $R$ and $z$ for $\dot{m}=0.1,~1,~10$ are shown in Figure 5. The heavy lines represent the surfaces of the disks. One end of the contours connects the disk surface and the other end tends to the equatorial plane. For the self-gravitating NDAF, it is noticed that $Q=1$ appears at the disk surface, corresponding to $R/R_{\rm g}\sim550$, $250$, and $35$ for the accretion rate $\dot m = 0.1,~1,~10$, and the region satisfied with $Q<1$ is more close to the central black hole for high accretion rates.

\subsection{Neutrino Luminosity}

The neutrino cooling rate $Q_\nu$ can be obtained according to the above calculations, thus the neutrino radiation luminosity $L_{\rm \nu}$ is expressed as
\beq
L_{\rm \nu}=2 \pi \int_{R_{\rm in}}^{R_{\rm out}} Q_{\rm \nu} R d R.
\eeq
In our calculations, the inner and outer edge of the disk are taken to be $R_{\rm in} = 3R_{\rm g}$ and $R_{\rm out} = 200R_{\rm g}$, respectively.

Figure 6 displays the neutrino luminosity for varying $\dot m$ from 0.1 to 10. We notice that the self-gravity has the limited effects on the neutrino luminosity of the NDAF. Compared with Case II, there exists an unremarkable increase in Case I only for the high accretion rate. The physical understanding is as follows. If the neutrino trapping process can be ignored \citep[e.g.,][]{Xue2013}, the higher density and temperature exist, the more neutrinos produced. The self-gravity plays such a role to partly enhance the conditions, especially for the outer region of the NDAF. However, the inner part of the NDAF is the major neutrino emission region \citep[e.g.,][]{Gu2006,Liu2007}, the self-gravity has little influence on the region as shown in the above figures, thus there was little variation of the luminosity, even for the high accretion rate. Moreover, the descriptions of microphysics are quite important for the NDAFs, especially for the structure and components in the outer region of the disk. In our previous work, we noticed that the descriptions of microphysics have little effect on the neutrino luminosity. In \citet{Xue2013} and \citet{Li2013}, we discussed the strict microphysics as far as possible and calculate the neutrino luminosity which is close to the results under the simple equation of state \citep[e.g.,][]{Popham1999,Di Matteo2002,Gu2006}. The main physical reason is as follows. Most neutrinos are launched from the inner region of the disk, and the main components in this region are the free baryons, which completely dominate the pressure. Thus the simple or complex descriptions of the microphysics have limited influence on the state of the inner region.

\section{Conclusions and discussion}

In this paper we have revisited the vertical structure of NDAFs in cylindrical coordinates by including the effects of the self-gravity. It is found that the significant changes of the structure appear in the outer part of the disk, especially for high accretion rates, and thus causes the neutrino luminosity slightly enhanced. Furthermore, the vertical distribution of the Toomre parameter is reviewed, which implies that the instability may occur in the outer region of the disk.

The criterion, $Q<1$, indicates that the disk is gravitationally unstable, which may cause two classes of possible behavior \citep[e.g.,][]{Perna2006}. Firstly, if the local cooling of the disk is rapid, the disk may fragment into two or more parts \citep[e.g.,][]{Nelson2000}. Since fragments form and the fall back timescale is long enough, the accretion processes will restart. This mechanism may explain the origin of late-time X-ray flares in GRBs \citep[e.g.,][]{Luo2013}, especially for short-duration GRBs corresponding to the high accretion rates in NDAFs. Secondly, the disk may evolve a quasi-steady spiral structure which can transfer angular momentum outward and mass inward. This mode can drive long-duration violent bursts if the disk mass is large enough \citep[e.g.,][]{Lodato2005}, which may be related to the origin of GRBs with extended emissions \citep[e.g.,][]{Liu2012b,Cao2014} or flares in long-duration GRBs.

X-ray flares are widely detected by \emph{Swift} and other telescopes \citep[e.g.,][]{Chincarini2007,Falcone2007}. The major concern is the relation between prompt emission and flares. For example, the lag-luminosity relation for X-ray flares has been investigated \citep{Margutti2010}, and it is similar to that of the prompt emission, as well as other statistical relations, which suggests that flares may share uniform origins with prompt emission. The flare models should be followed this origination principle. In turn, the principle is the only constraint on the theoretical models.

Except for the gravitational instability, several other mechanisms have been proposed to explain the episodic X-ray flares in GRBs, such as fragmentation of a rapidly rotating core \citep{King2005}, a magnetic switch of the accretion process \citep{Proga2006}, differential rotation in a post-merger millisecond pulsar \citep{Dai2006}, transition of the accretion modes \citep{Lazzati2008}, outflow caused by the extreme mass accretion rates of NDAFs \citep{Liu2008}, He-synthesis-driven wind \citep{Lee2009}, jet precession \citep{Liu2010b,Hou2014b}, dynamical instability in the jet \citep{Lazzati2011}, episodic jet produced by the magnetohydrodynamic mechanism from the accretion disk \citep{Yuan2012}, and so on. Thus, more information, which is from the future multi-band observations and the detections on the polarization and gravitational waves on the GRBs and their flares, should be given in order to identify these models.

\begin{acknowledgements}
We thank the anonymous referee for very useful suggestions and comments. This work was supported by the National Basic Research Program of China (973 Program) under grant 2014CB845800, the National Natural Science Foundation of China under grants 11103015, 11163003, 11222328, 11233006, 11333004, and U1331101, the CAS Open Research Program of Key Laboratory for the Structure and Evolution of Celestial Objects under grant OP201305, and the Natural Science Foundation of Fujian Province of China under grant 2012J01026.
\end{acknowledgements}

\begin{figure}
\centering
\includegraphics[angle=0,scale=0.9]{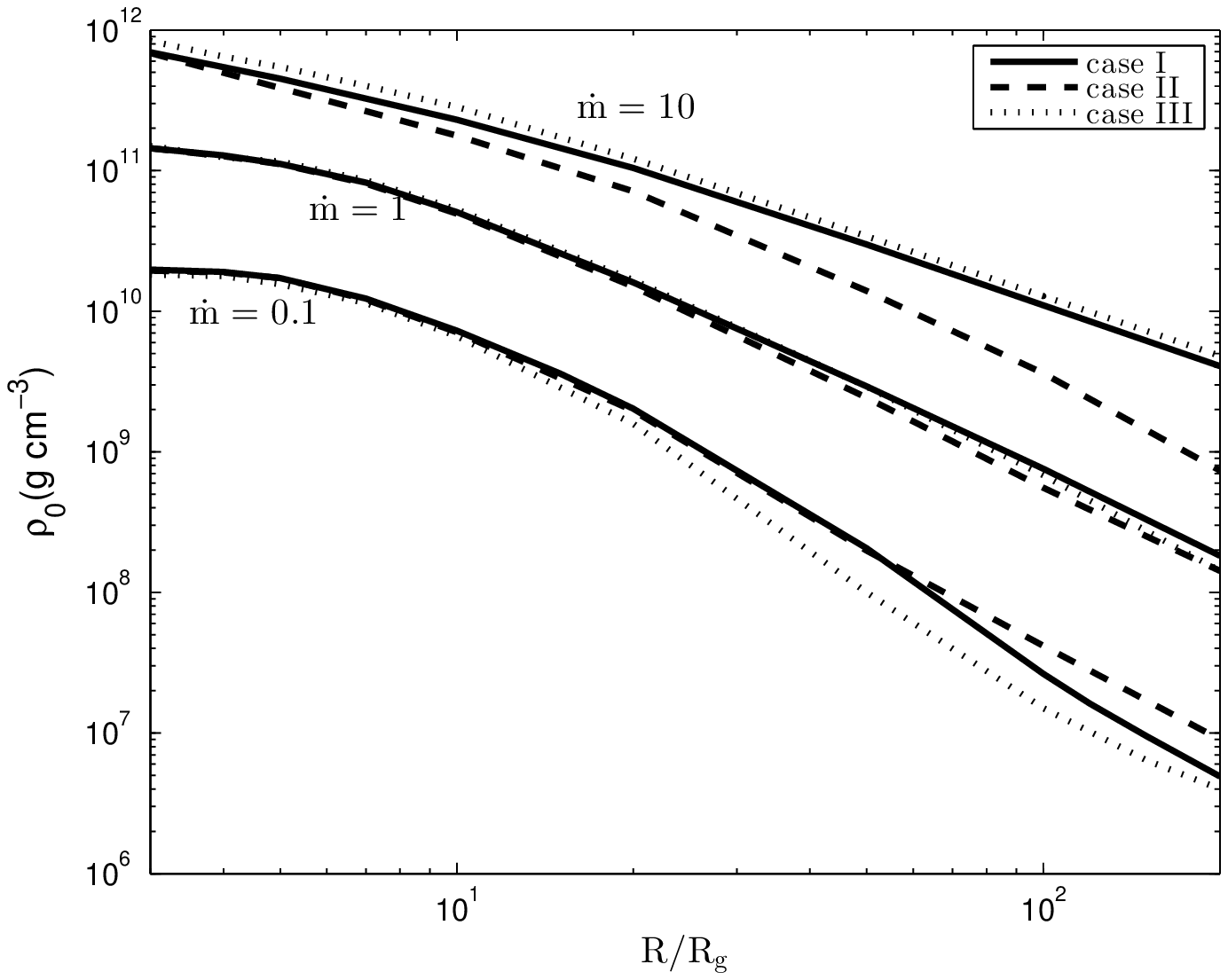}
\caption{Variations of the mass density on the equatorial plane $\rho_0$ with radius $R$ for $\dot{m}=0.1,~1,~10$ ($\dot M=\dot{m} M_{\odot}~\rm s^{-1}$). Case I, II, and III are described by the solid, dashed, and dotted lines, respectively (similarly hereafter).}
\end{figure}

\clearpage

\begin{figure}
\centering
\includegraphics[angle=0,scale=0.8]{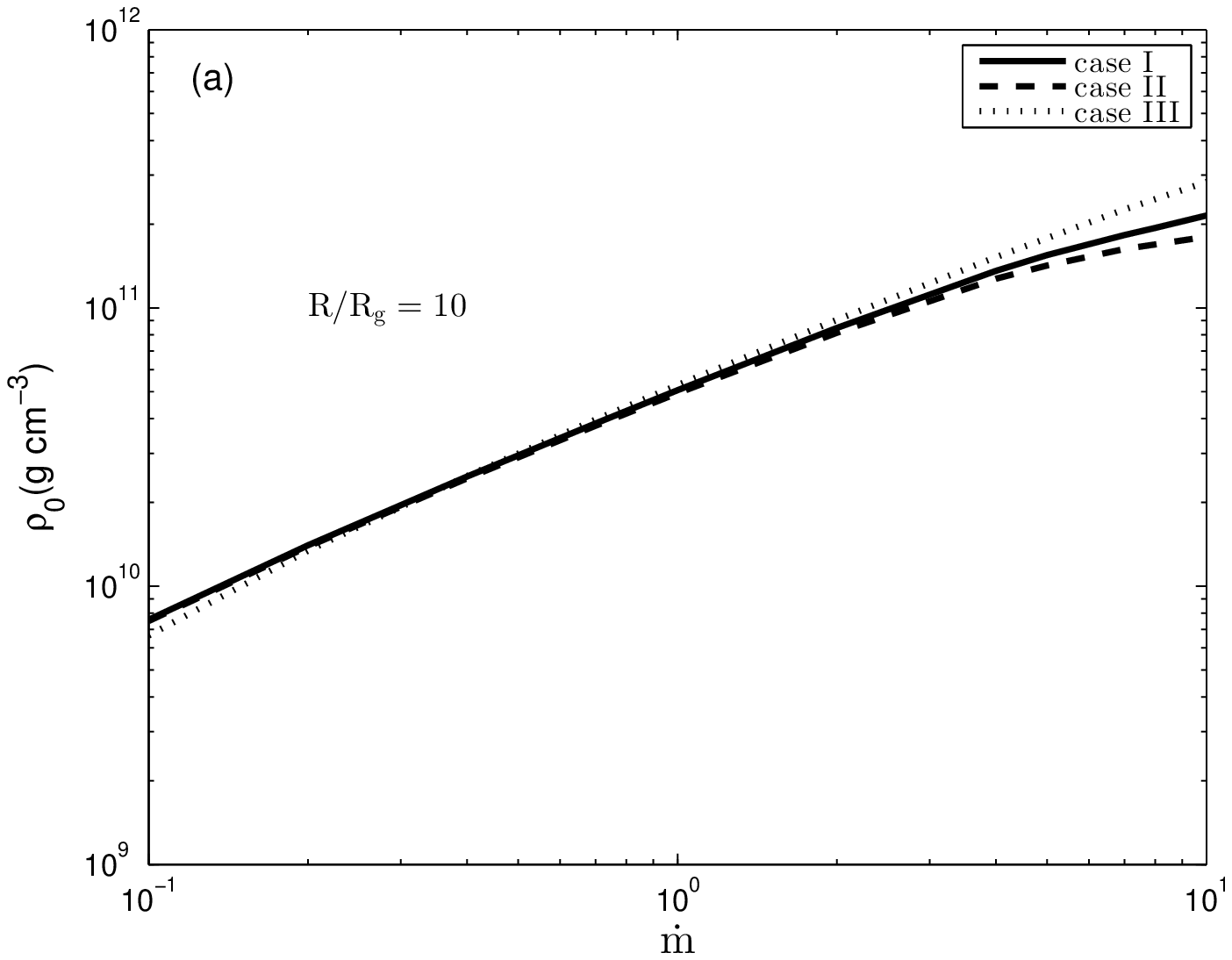}
\includegraphics[angle=0,scale=0.8]{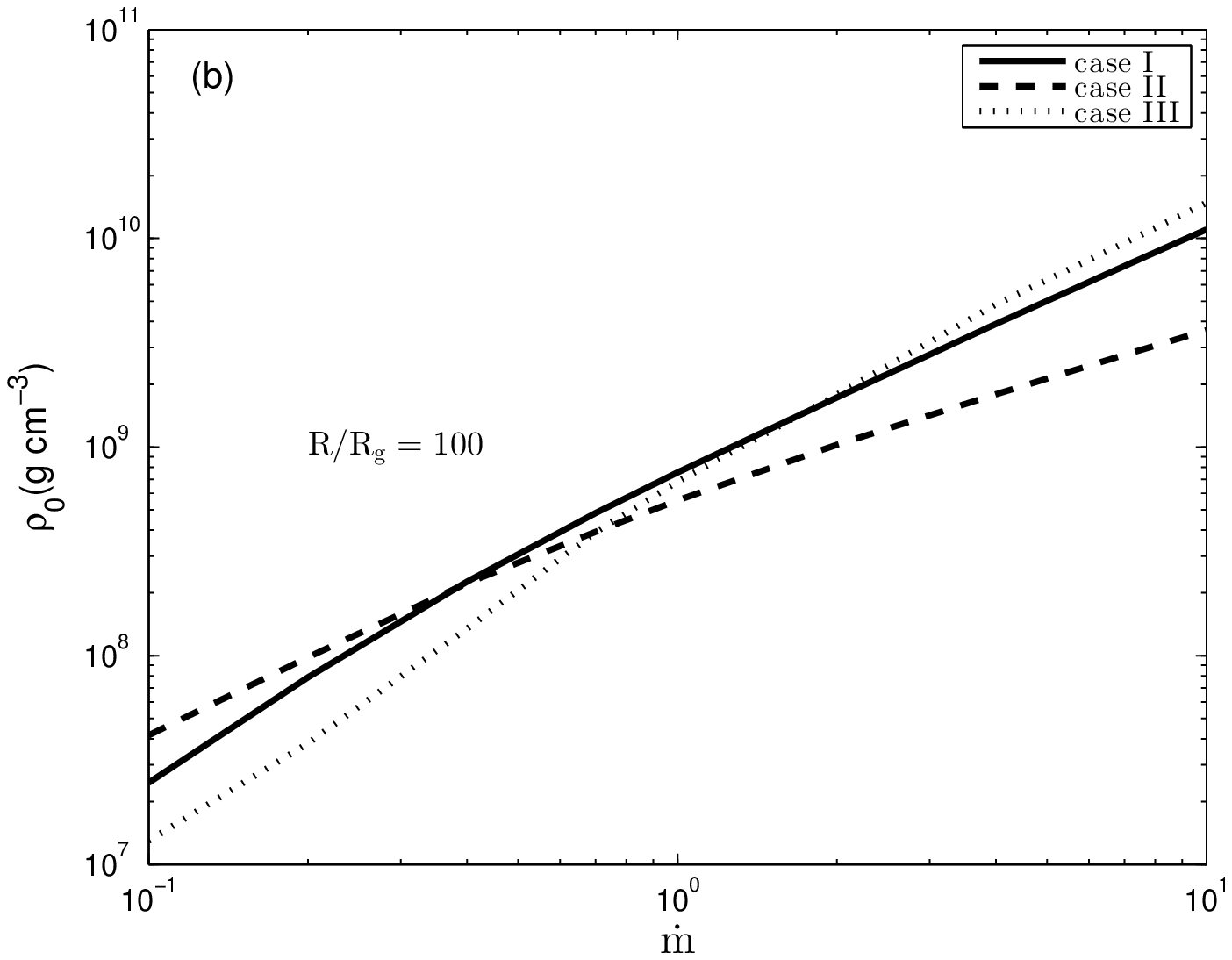}
\caption{Variations of $\rho_0$ with accretion rate $\dot m$ at $R=$ 10$R_{\rm g}$ and 100$R_{\rm g}$.}
\end{figure}

\clearpage

\begin{figure}
\centering
\includegraphics[angle=0,scale=0.6]{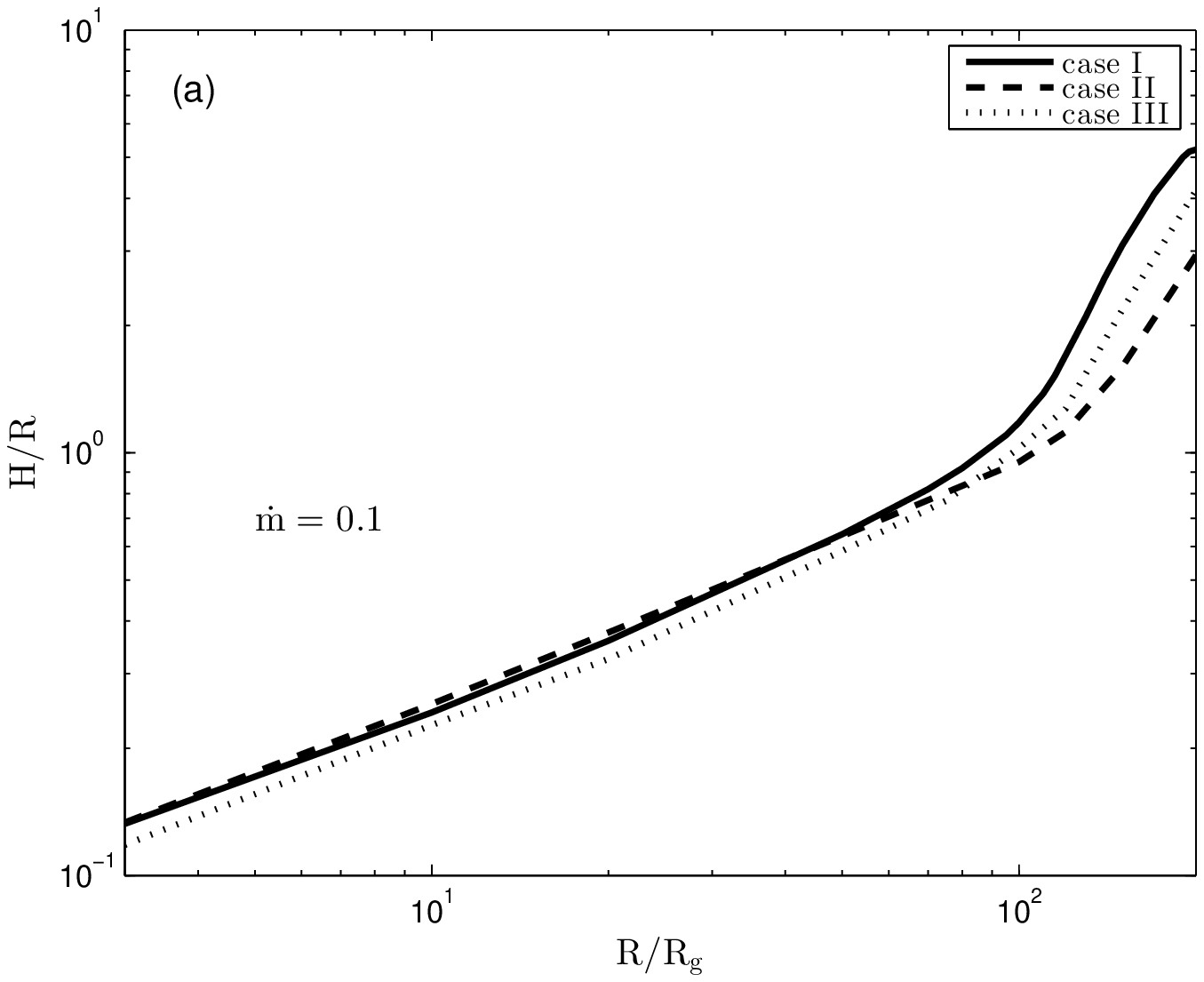}
\includegraphics[angle=0,scale=0.6]{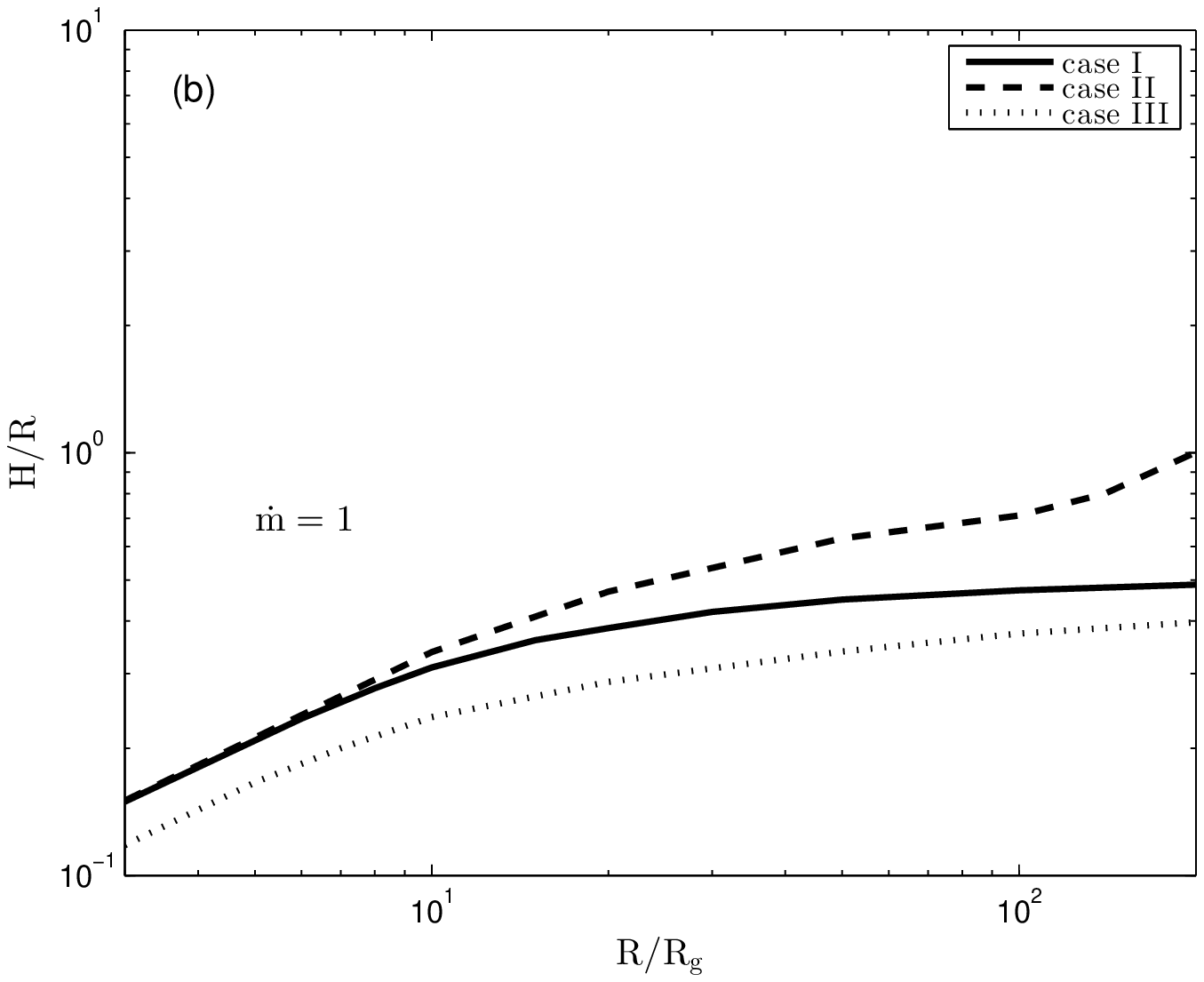}
\includegraphics[angle=0,scale=0.6]{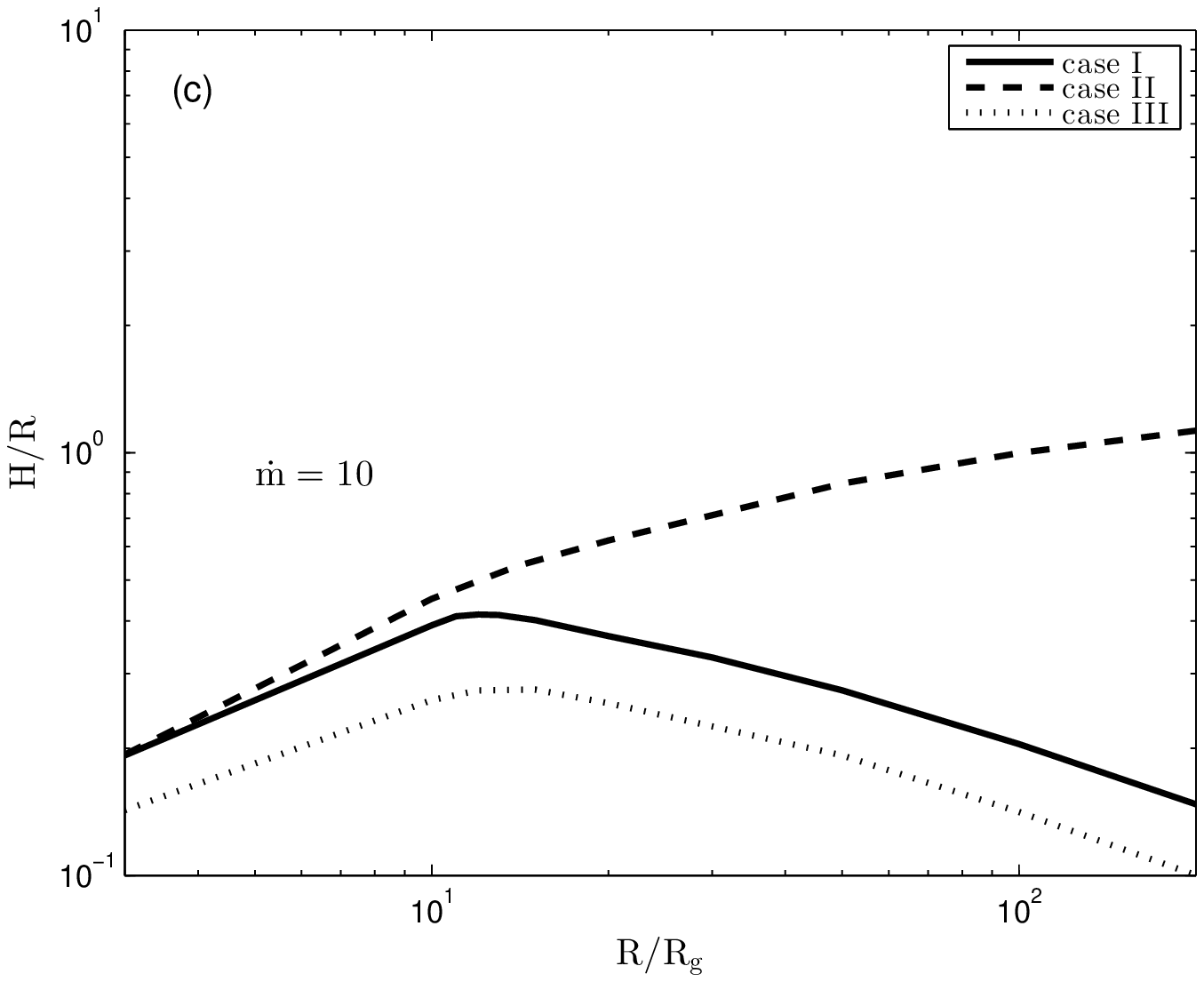}
\caption{Relative thickness $H/R$ as a function of $R$ for $\dot m=0.1,~1,~10$.}
\end{figure}

\clearpage

\begin{figure}
\centering
\includegraphics[angle=0,scale=0.8]{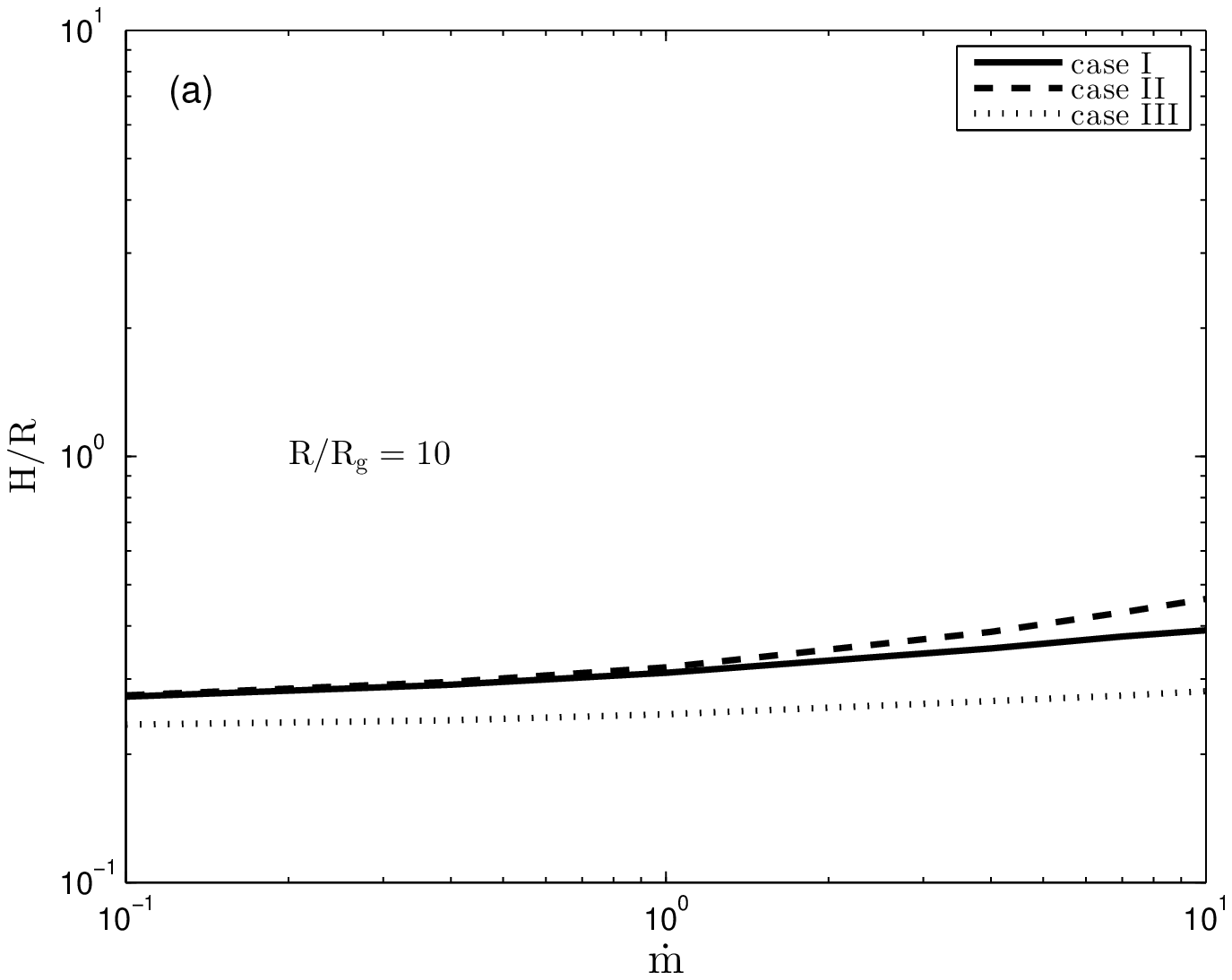}
\includegraphics[angle=0,scale=0.8]{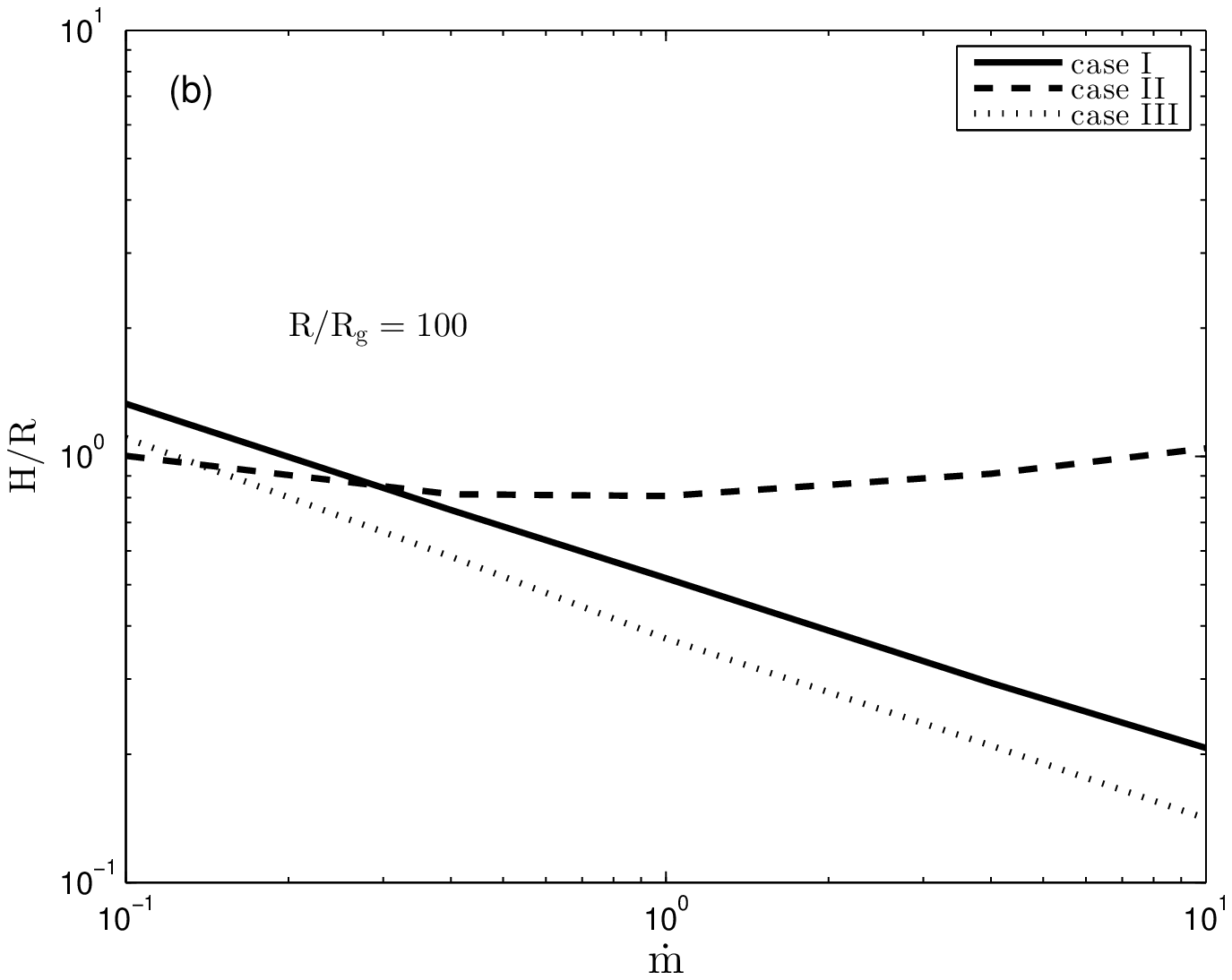}
\caption{Relative thickness $H/R$ for varying accretion rate $\dot m$ from 0.1 to 10 for fixed radii 10$R_{\rm g}$ and 100$R_{\rm g}$.}
\end{figure}

\clearpage

\begin{figure}
\centering
\includegraphics[angle=0,scale=0.6]{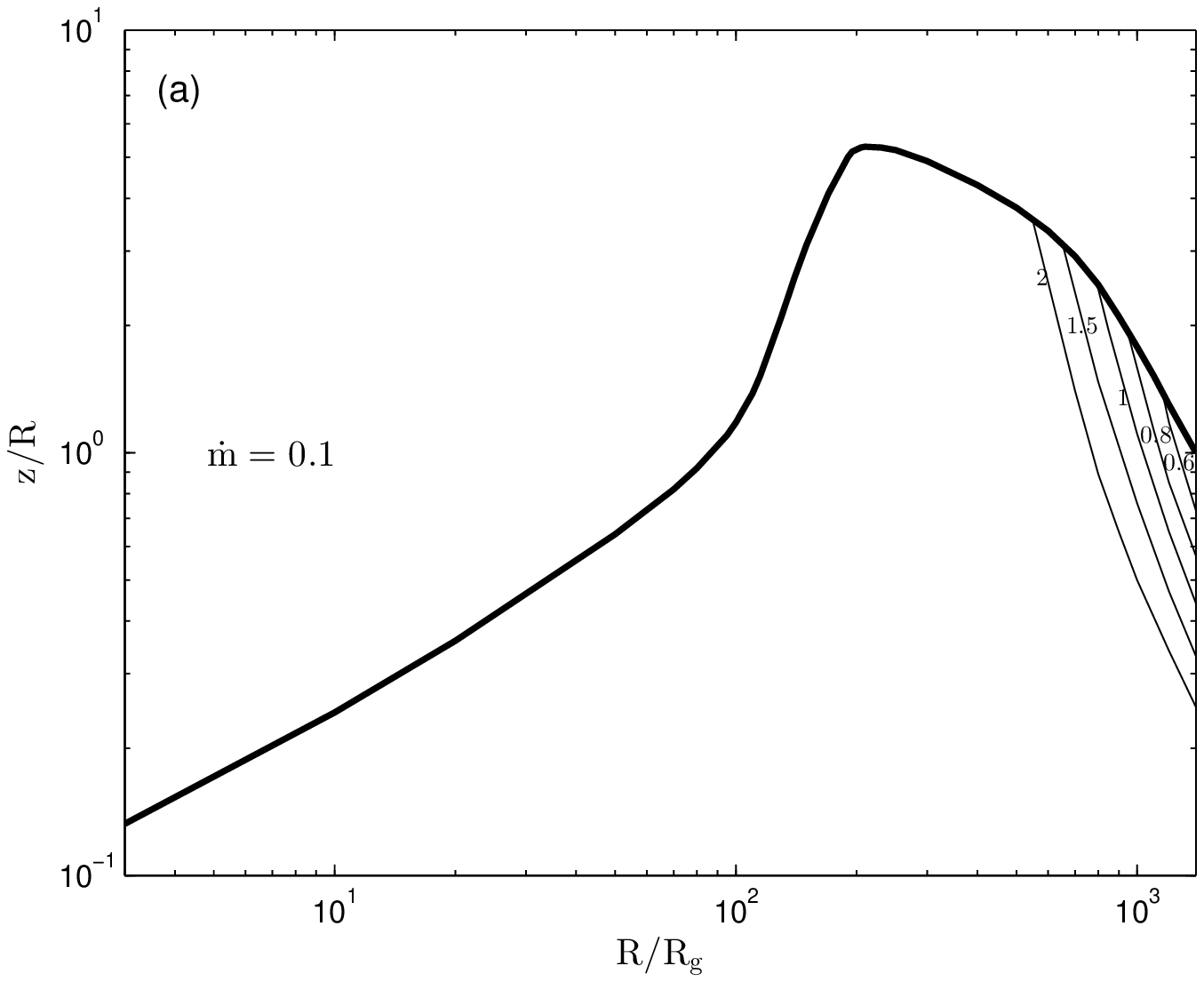}
\includegraphics[angle=0,scale=0.6]{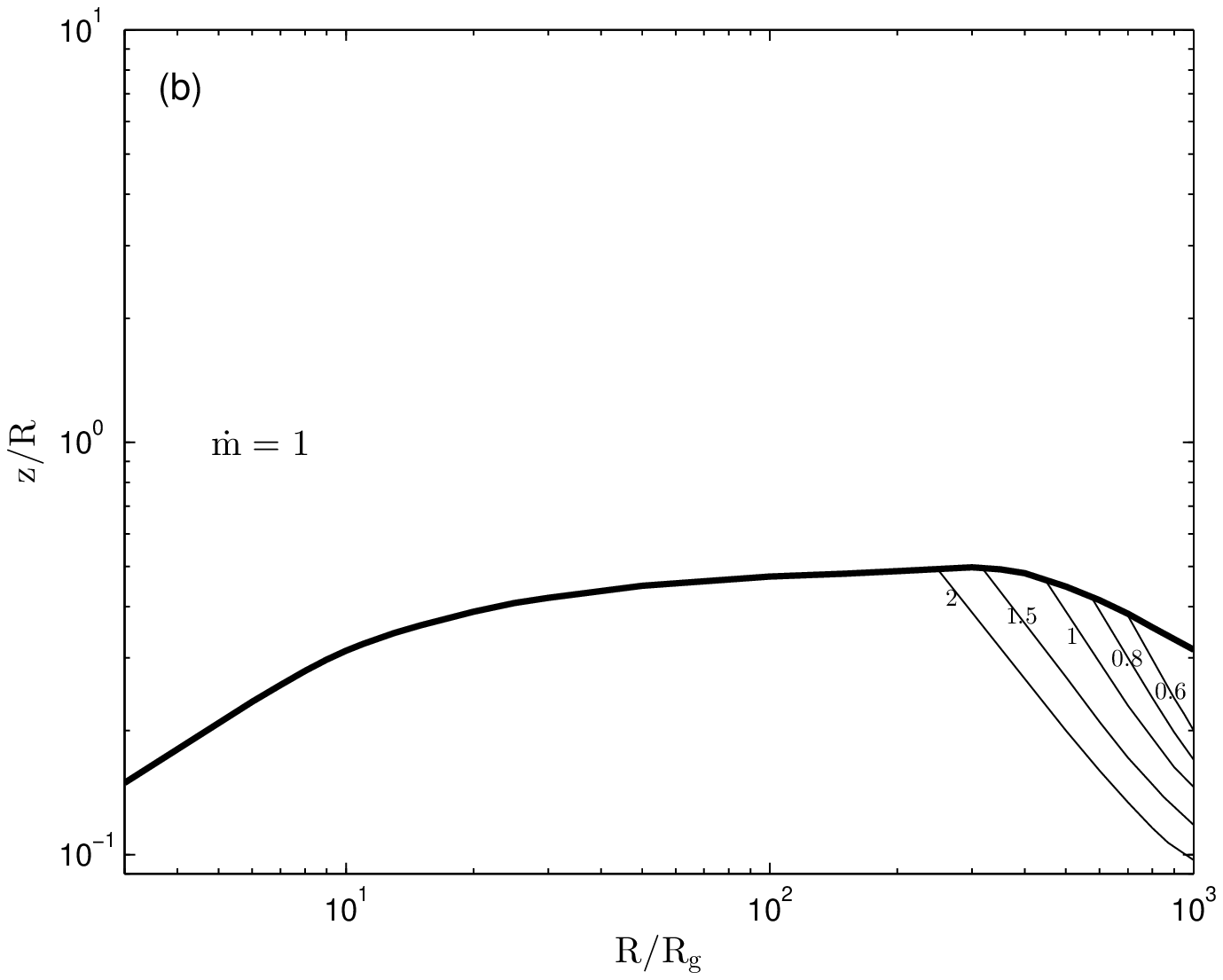}
\includegraphics[angle=0,scale=0.6]{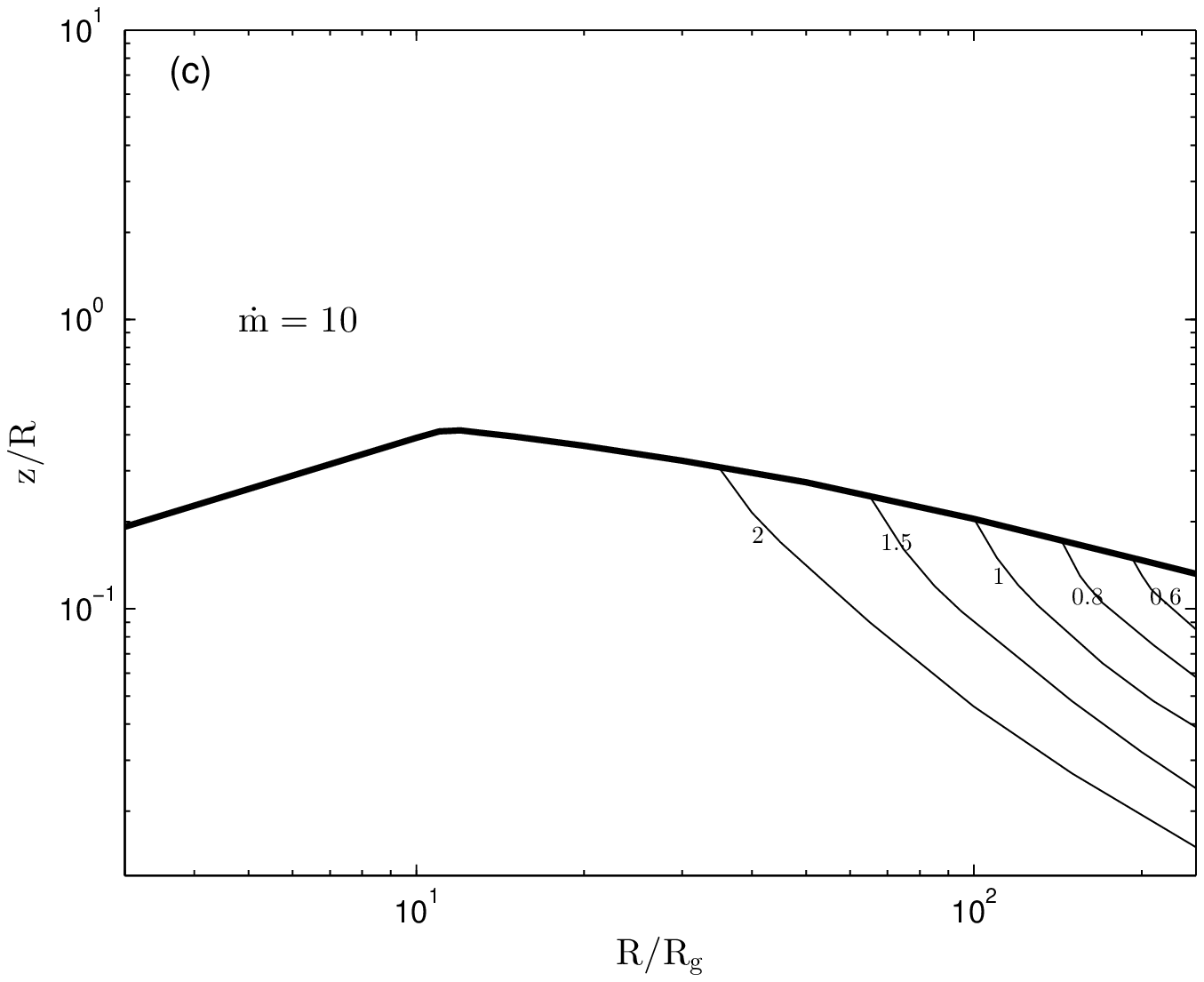}
\caption{Contours of Toomre parameter with cylindrical coordinates $R$ and $z$ for $\dot{m}=0.1,~1,~10$.}
\end{figure}

\clearpage

\begin{figure}
\centering
\includegraphics[angle=0,scale=0.9]{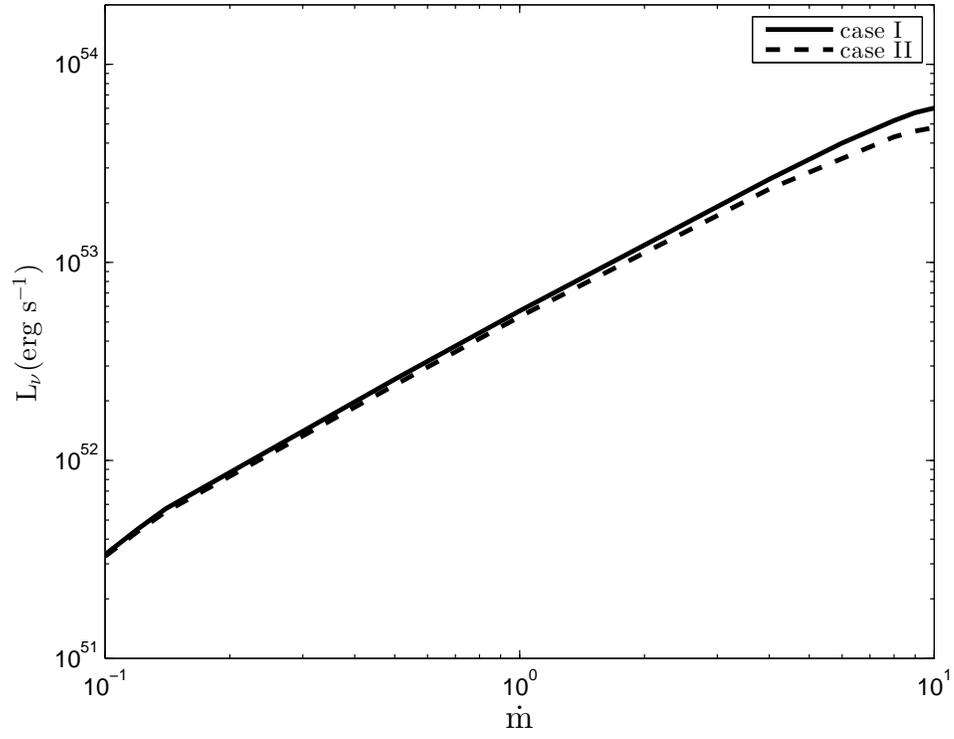}
\caption{Neutrino luminosity $L_\nu$ for varying $\dot m$ from 0.1 to 10.}
\end{figure}

\clearpage

\end{document}